\documentclass[%
reprint,
onecolumn,
 amsmath,amssymb,
 aps,
]{revtex4-1}

\usepackage{graphicx}
\usepackage{dcolumn}
\usepackage{bm}
\usepackage{epsfig}
\usepackage{tabularx}

\begin{document}

\preprint{APS/123-QED}

\title{Three-dimensional cascaded lattice Boltzmann method: improved implementation and consistent forcing scheme}

\author{ Linlin Fei$^{1}$, Kai Hong Luo$^{1,2}$\footnote{Corresponding author: K.Luo@ucl.ac.uk}, Qing Li$^{3}$}
\affiliation{$^1$ Center for Combustion Energy; Key laboratory for Thermal Science and Power Engineering of Ministry of Education, Department of Energy and Power Engineering, Tsinghua University, Beijing 100084, China \\
	$^2$ Department of Mechanical Engineering, University College London, Torrington Place, London WC1E 7JE, UK\\
	$^3$ School of Energy Science and Engineering, Central South University, Changsha 410083, China	
}

\date{\today}

\begin{abstract}
Cascaded or central-moment-based lattice Boltzmann method (CLBM) proposed in [Geier \textit{et al.}, Phys. Rev. E \textbf{63}, 066705 (2006)] possesses very good numerical stability. However, two constraints exist in three-dimensional (3D) CLBM simulations. Firstly, the conventional implementation for 3D CLBM involves cumbersome operations and requires much higher computational cost compared to the single-relaxation-time (SRT) LBM. Secondly, it is a challenge to accurately incorporate a general force field into the 3D CLBM.  In this paper, we present an improved method to implement CLBM in 3D. The main strategy is to adopt a simplified central moment set, and carry out the central-moment-based collision operator based on a general multi-relaxation-time (GMRT) framework. Next, the recently proposed 
consistent forcing scheme in CLBM [L. Fei and K. H. Luo, Phys. Rev. E \textbf{96}, 053307 (2017)] is extended to incorporate a general force field into 3D CLBM.
Compared with the recently developed non-orthogonal CLBM [A. D. Rosis, Phys. Rev. E \textbf{95}, 013310 (2017)], our  implementation is proved to  reduce the computational cost significantly. The inconsistency of adopting the discrete equilibrium distribution functions (EDFs) in the non-orthogonal CLBM is revealed and discussed.
The 3D CLBM developed here in conjunction with the consistent forcing scheme is verified through numerical simulations of several canonical force-driven flows, highlighting very good properties in terms of accuracy, convergence and consistency with the nonslip rule. Finally, the techniques developed here for 3D CLBM can be applied to make the implementation and execution of 3D MRT-LBM much more efficient.

\begin{description}
	\item[PACS numbers]
	47.11.-j, 05.20.Dd
\end{description}
\end{abstract}                             

\maketitle


\section{Introduction}
As a mesoscopic numerical method based on the kinetic theory, the lattice Boltzmann method (LBM) has gained remarkable success for the simulation of complex fluid flows and beyond during the last three decades or so \cite{qian1995recent,chen1998lattice,succi2001lattice,shan1993lattice,gan2011lattice,guo2013lattice,li2015lattice,gan2015discrete,xu2015three,li2016lattice,xu2016three,lin2016double,Gong2017Wetting,lin2017multi,gong2017thermal}. Different from the conventional computational fluid dynamics (CFD) methods where the macroscopic governing equations are solved numerically, LBM solves a discrete Boltzmann equation which is designed to reproduce the Navier-Stokes (N-S) equations in the macroscopic limit. In the LBM simulation, the fluid is usually represented by populations of fictitious particles colliding locally and streaming to adjacent nodes along the links of a specified lattice. The scale-bridging nature of LBM allows its natural incorporation of microscopic and/or mesoscopic physics, while the highly efficient “collision-streaming” algorithm makes it affordable computationally \cite{li2016lattice}.

In the practical implementation, the simplest approach to represent  “particles colliding” process is to relax all the distribution functions (DFs) to their local equilibria at an identical rate, which is known as the
single-relaxation-time (SRT) model \cite{qian1992lattice}. While the SRT-LBM is successful in many fluid flows, it may encounter numerical instability for flows with relatively low viscosities\cite{d1994generalized,lallemand2000theory}, as well as inaccuracy in implementing the boundary  conditions\cite{ginzburg2003multireflection,pan2006evaluation}. Compared with the SRT model, the multiple-relaxation-time (MRT) model, where the collision step is carried out in the raw moment space,
is able to enhance the stability of LBM by carefully separating the time scales among the kinetic modes \cite{d1994generalized,lallemand2000theory,d2002multiple}. In addition, the MRT-LBM can also improve the numerical accuracy for non-slip boundary conditions by adopting a so-called “magic” parameter \cite{ginzburg2003multireflection,pan2006evaluation}. In 2006, a cascaded collision operator was proposed by Geier \textit{et al.} \cite{geier2006cascaded}, where the collision step is implemented in the central moment space. To get the higher-order post-collision central moments, the lower-order ones are needed, which implies a “from the lowest order to the highest order”，i.e., cascaded operation procedure. The cascaded lattice Boltzmann method (CLBM) increases the numerical stability significantly, which is also essentially due to the removal of the “ghost modes”. Besides, relaxation in a co-moving frame of reference, i.e., in terms of central moments, allows a natural setting to achieve better Galilean invariance, compared with in the frame at rest, for a specified discrete velocity set \cite{geier2006cascaded}. More comparisons and discussions between relaxations in the raw moment and central moment spaces can be found in \cite{geier2006cascaded,premnath2009incorporating,premnath2011three,chen2014recovery,geier2017parametrization}. 

More recently, many studies have been carried out to improve the cascaded collision model and  apply the CLBM to practical applications. In \cite{Asinari2008Generalized}, Asinari argued that CLBM essentially consists in using a generalized local 
equilibrium in the frame at rest. In addition, a multiphase CLBM has been developed to simulate multiphase flows by Lycett-Brown and Luo \cite{lycett2014multiphase,lycett2014binary}.  They further extended the model with an improved forcing scheme and achieved binary collision simulations at high parameters \cite{lycett2016cascaded}. Moreover, a thermal cascaded LBM (TCLBM) has been proposed by the present authors to simulate low-Mach compressible thermal flows \cite{fei2016thermal}, and several different CLBMs have been developed later for incompressible thermal flows \cite{fei2018modeling,fei2018cascaded,shah2017cascaded,kumar2017numerical,sharma2017new}. Finally, CLBM has also been extended to simulate shallow water equations\cite{de2017centralshallow}, moving boundary problems \cite{falagkaris2017proteus}, as well as stationary flows with a preconditioning method \cite{hajabdollahi2017improving}.

Although the CLBM possesses very good numerical stability and has achieved success for a series of complex flows, two critical problems still exist in the three-dimensional (3D) simulations. Firstly, the practical implementation for 3D CLBM in the original work \cite{geier2006cascaded} involves a lot of
cumbersome notations and the computational cost is much higher than the SRT-LBM. Even if some efforts have been made \cite{premnath2011three,lycett2014binary,de2017nonorthogonal}, it is still quite difficult to handle the expressions compared with the SRT-LBM. For example, a naive implementation of the most recent non-orthogonal CLBM \cite{de2017nonorthogonal} needs a CPU time that is 13 to 14 times larger than that by the SRT-LBM. Secondly, an accurate and easy-to-implement forcing scheme is needed to incorporate an external or internal force field into the 3D CLBM. In 2009, Premnath \textit{et al.} proposed a forcing scheme for CLBM by method of central moments \cite{premnath2009incorporating}, which was then extended to 3D model in \cite{premnath2011three}. In \cite{lycett2014multiphase,lycett2014binary,lycett2016cascaded}, the SRT-style 
forcing scheme was used in the CLBM. Besides, an alternative forcing scheme based on a discrete equilibrium has been developed by De Rosis \cite{de2017alternative}.
However, 
as analyzed by the present authors \cite{fei2017consistent}, the forcing schemes in \cite{premnath2009incorporating,lycett2014multiphase,de2017alternative} will lead to some inconsistences for flows with a general force field. Based on a general multi-relaxation-time (GMRT) framework, a consistent forcing scheme in CLBM was proposed in \cite{fei2017consistent}. The consistent forcing scheme can degrade into the widely used forcing schemes in the MRT-LBM \cite{Guo2008Analysis} and SRT-LBM \cite{guo2002discrete} under certain conditions and shows great advantages over previous forcing schemes in terms of accuracy, isotropy and consistency with the nonslip rule. Therefore, the present paper aims to simplify the implementation and reduce the computational cost for 3D CLBM. In the meantime, the consistent forcing scheme is extended to incorporate a general force field for 3D flows. 

The rest of this paper is organized as follows. Section \ref{sec.2} presents the simplified implementation and the consistent forcing scheme for 3D CLBM. Numerical verifications are carried out in Sec. \ref{sec.3}. Finally, concluding remarks are given in Sec. \ref{sec.4}.
\section{SIMPLIFIED 3D CLBM WITH CONSISTENT FORCING SCHEME}\label{sec.2}
The simplified implementation is based on the GMRT framework and adopts a new central moment set. Firstly, the GMRT framework with the consistent forcing scheme is introduced briefly. Then, the choices of the central moment set, central moment equilibria, and forcing terms in the central moment space are given in detail.
\subsection{GMRT framework}\label{sec.2a}
In the present work, we focus on the standard D3Q27 discrete velocity model (DVM). However, it should be noted that the procedures shown in this work are not limited to the specified DVM, and can be extended to other DVMs readily. For example, a D3Q19 CLBM can be directly extracted from the D3Q27 CLBM (see in the Appendix \ref{sec.5a}). The lattice speed  $ c=\Delta{x}/\Delta {t}=1 $ and the lattice sound speed $ c_{s}=1/\sqrt{3} $ are adopted, in which $ \Delta{x} $ and $ \Delta {t} $ are the lattice spacing and time step. The discrete velocities $ {{\bf{e}}_i} = \left[ {\left| {{e_{ix}}} \right\rangle ,\left| {{e_{iy}}} \right\rangle ,\left| {{e_{iz}}} \right\rangle } \right] $ are defined as 
\begin{equation}\label{e1}
\begin{array}{l}
\left| {{e_{ix}}} \right\rangle  = {[0,1, - 1,0,0,0,0,1, - 1,1, - 1,1, - 1,1, - 1,0,0,0,0,1, - 1,1, - 1,1, - 1,1, - 1]^ \top }, \\ 
\left| {{e_{iy}}} \right\rangle  = {[0,0,0,1, - 1,0,0,1,1, - 1, - 1,0,0,0,0,1, - 1,1, - 1,1,1, - 1, - 1,1,1, - 1, - 1]^ \top }, \\ 
\left| {{e_{i{\rm{z}}}}} \right\rangle  = {[0,0,0,0,0,1, - 1,0,0,0,0,1,1, - 1, - 1,1,1, - 1, - 1,1,1,1,1, - 1, - 1, - 1, - 1]^ \top }. \\ 
\end{array}
\end{equation}
where $ i = 0,1,...,26 $, ${\left|  \cdot  \right\rangle }$ denotes a 27-dimensional column vector, and the superscript $ \top $ denotes the transposition.  

We first define the raw and central moments of the discrete distribution functions (DFs) ${{f_i}}$,
\begin{equation}\label{e2}
\begin{array}{l}
{k_{mnp}} = \left\langle {{f_i}\left| {e_{ix}^me_{iy}^ne_{iz}^p} \right.} \right\rangle ,~~{\rm{ }}{{\tilde k}_{mnp}} = \left\langle {{f_i}\left| {{{({e_{ix}} - {u_x})}^m}{{({e_{iy}} - {u_y})}^n}{{({e_{iy}} - {u_z})}^p}} \right.} \right\rangle  , \\ 
\end{array}
\end{equation}
where $ m $, $ n $ and $ p $ are integers, and $ {u_x} $, $ {u_y} $ and $ {u_z} $ are velocity components in $ x $, $ y $ and $ z $ directions, respectively. The equilibrium values $ k_{_{{mnp}}}^{eq} $ and $ \tilde k_{{mnp}}^{eq}$ are defined
analogously by replacing ${{f_i}}$ with the discrete equilibrium distribution functions (EDFs) $ {f_i^{eq}} $. To construct the central-moment-based collision operator, a raw moment set $\left| {{T_i}} \right\rangle$ and the corresponding central moment set $\left| {{{\tilde T}_i}} \right\rangle $ are needed,
\begin{equation}\label{e3}
\left| {{T_i}} \right\rangle  = {\left[ {{T_0},{T_1},...,{T_{26}}} \right]^ \top },
~~\left| {{{\tilde T}_i}} \right\rangle  = {\left[ {{{\tilde T}_0},{{\tilde T}_1},...,{{\tilde T}_{26}}} \right]^ \top },
\end{equation}
where the elements in $\left| {{T_i}} \right\rangle $ and $\left| {{{\tilde T}_i}} \right\rangle $ are combinations of ${k_{mnp}}$ and ${{\tilde k}_{mnp}}$ in the ascending order of ($m+n+p$), respectively. According to Eq. (\ref{e2}),
the transformation from the discrete DFs to their raw moments can be performed through a transformation matrix $ {\bf{M}} $, and the shift from the raw moments to central moments can be performed through a shift matrix $ {\bf{N}} $,
\begin{equation}\label{e4}
\begin{array}{l}
\left| {{{\tilde T}_i}} \right\rangle  = {\bf{N}}\left| {{T_i}} \right\rangle  = {\bf{NM}}\left| {{f_i}} \right\rangle 
\end{array}
\end{equation}
The explicit forms for $ {\bf{M}}$ and $ {\bf{N}}$ depend on the raw moment set and the corresponding central moment set, which will be discussed in the next subsection.

In the implementation of CLBM, the collision step is firstly executed in the central moment space. To be consistent with the central-moment-based collision operator, an external or internal force field ${\bf{F}} = [{F_x},{F_y},{F_z}]$ should be added by means of central moments \cite{premnath2009incorporating}. By using a transformation method to incorporate forcing terms with a second-order trapezoidal scheme, the explicit form of collision step can be written as \cite{fei2017consistent},
\begin{equation}\label{e5}
\left| {\tilde T_i^*} \right\rangle  = ({\bf{I}} - {\bf{S}})\left| {{{\tilde T}_i}} \right\rangle  + {\bf{S}}\left| {\tilde T_i^{eq}} \right\rangle  + ({\bf{I}} - {\bf{S}}/2)\left| {{C_i}} \right\rangle  = ({\bf{I}} - {\bf{S}}){\bf{NM}}\left| {{f_i}} \right\rangle  + {\bf{SNM}}\left| {f_i^{eq}} \right\rangle  + ({\bf{I}} - {\bf{S}}/2){\bf{NM}}\left| {{R_i}} \right\rangle, 
\end{equation}
where  ${\bf{I}} $ is a unit matrix, ${\bf{S}}$ is the relaxation matrix, and ${{C_i}} $ and ${{R_i}}$ are forcing terms in central moment space and discrete velocity space, respectively. According to the assumption by He \textit{et al}. \cite{he1998novel},  ${{R_i}}$ can be given by
\begin{equation}\label{e6}
{R_i} = \frac{{\bf{F}}}{\rho }\frac{{\left( {{{\bf{e}}_i} - {\bf{u}}} \right)}}{{c_s^2}}f_i^{eq}.
\end{equation}
Due to the definitions of the transformation and shift matrices, both of them are invertible (explicit expressions for ${{\bf{M}}^{ - 1}}$ and ${{\bf{N}}^{ - 1}}$
can be easily obtained by software like MATLAB). The post-collision discrete DFs can be reconstructed by
\begin{equation}\label{e7}
\begin{array}{l}
\left| {f_i^*} \right\rangle  = {{\bf{M}}^{ - 1}}\left| {T_i^*} \right\rangle, ~
\left| {T_i^*} \right\rangle {\rm{ = }}{{\bf{N}}^{ - 1}}\left| {\tilde T_i^*} \right\rangle 
\end{array}
\end{equation}

In the  streaming step, the post-collision discrete DFs in space $ \bf{x} $ stream to their neighbors $({\bf{x}} + {{\bf{e}}_i}\Delta t)$ along the characteristic lines as usual \cite{li2016lattice,fei2017consistent},
\begin{equation}\label{e8}
{f_i}(\textbf{x} + {\textbf{e}_i}\Delta t,t + \Delta t) = f_i^*(\textbf{x},t).
\end{equation}
Then, the fluid density and velocity, $ \rho $ and ${\bf{u}} = [{u_x},{u_y},{u_z}]$, are updated by,
\begin{equation}\label{e9}
\rho  = \sum\nolimits_i {{f_i}} ,~~~\rho {\bf{u}} = \sum\nolimits_i {{f_i}} {{\bf{e}}_i} + \Delta t{\bf{F}}/2.
\end{equation}

From the above, it can be shown that when the shift matrix $ \textbf{N} $ is a unit matrix the CLBM degrades into an MRT-LBM on the specified raw moment set $\left| {{T_i}} \right\rangle $, and when all the relaxation parameters in the matrix $ \textbf{S} $ are equal to one another the CLBM degrades into an SRT-LBM. Thus we proclaim the above framework as a GMRT framework \cite{fei2017consistent}. It can be also shown that the above forcing scheme can
degrade into the MRT version and SRT version of the widely used forcing scheme by  Guo \textit{et al.} \cite{guo2002discrete} under certain conditions. Thus it is named as a consistent scheme \cite{fei2017consistent}.

\subsection{Central moment set, equilibria, and forcing terms}\label{sec.2b}
In this subsection, we firstly discuss the central moment set, which 
is an important step to construct the CLBM. As discussed in \cite{premnath2011three,lycett2014binary}, the  conserved moments ($ {\tilde k_{000}} $, $ {\tilde k_{100}} $, $ {\tilde k_{010}} $, $ {\tilde k_{001}} $) should be considered to represent the mass and 
momentum conservations, the second-order moments are chosen such that it allows correct representation of the momentum flux in the hydrodynamic equations, while the rest moments can be chosen order by order under moments-independence constraint for a specified DVM. Premnath \textit{et al.} adopted an orthogonal central moment set \cite{premnath2011three}, 
\begin{equation}\label{e10}
\begin{array}{l}
\left| {{{\tilde T}_i}} \right\rangle  = [{{\tilde k}_{000}},{{\tilde k}_{100}},{{\tilde k}_{010}},{{\tilde k}_{001}},{{\tilde k}_{110}},{{\tilde k}_{101}},{{\tilde k}_{011}},{{\tilde k}_{200}} - {{\tilde k}_{020}},({{\tilde k}_{200}} + {{\tilde k}_{020}} + {{\tilde k}_{002}}) - 3{{\tilde k}_{002}}, \\ 
({{\tilde k}_{200}} + {{\tilde k}_{020}} + {{\tilde k}_{002}}) - 2{{\tilde k}_{000}},3({{\tilde k}_{120}} + {{\tilde k}_{102}}) - 4{{\tilde k}_{100}},3({{\tilde k}_{210}} + {{\tilde k}_{012}}) - 4{{\tilde k}_{010}}, \\ 
3({{\tilde k}_{201}} + {{\tilde k}_{021}}) - 4{{\tilde k}_{001}},{{\tilde k}_{120}} - {{\tilde k}_{102}},{{\tilde k}_{210}} - {{\tilde k}_{012}},{{\tilde k}_{201}} - {{\tilde k}_{021}},{{\tilde k}_{111}}, \\ 
3({{\tilde k}_{220}} + {{\tilde k}_{202}} + {{\tilde k}_{022}}) - 4({{\tilde k}_{200}} + {{\tilde k}_{020}} + {{\tilde k}_{002}}) + 4{{\tilde k}_{000}},3({{\tilde k}_{220}} + {{\tilde k}_{202}} - 2{{\tilde k}_{022}}) \\ 
- 2(2{{\tilde k}_{200}} - {{\tilde k}_{020}} - {{\tilde k}_{002}}),3({{\tilde k}_{220}} - {{\tilde k}_{202}}) - 2(2{{\tilde k}_{020}} - {{\tilde k}_{002}}),3{{\tilde k}_{211}} - 2{{\tilde k}_{011}}, \\ 
3{{\tilde k}_{121}} - 2{{\tilde k}_{101}},3{{\tilde k}_{112}} - 2{{\tilde k}_{110}},9{{\tilde k}_{122}} - 6({{\tilde k}_{120}} + {{\tilde k}_{102}}) + 4{{\tilde k}_{100}},9{{\tilde k}_{212}} - 6({{\tilde k}_{210}} + {{\tilde k}_{012}}) + 4{{\tilde k}_{010}}, \\ 
9{{\tilde k}_{221}} - 6({{\tilde k}_{201}} + {{\tilde k}_{021}}) + 4{{\tilde k}_{001}},27{{\tilde k}_{222}} - 18({{\tilde k}_{220}} + {{\tilde k}_{202}} + {{\tilde k}_{022}}) + 12({{\tilde k}_{200}} + {{\tilde k}_{020}} + {{\tilde k}_{002}}) - 8{{\tilde k}_{000}}{]^ \top }. \\ 
\end{array}
\end{equation}
According to the binomial theorem, a central moment can be expressed by raw moments from the lowest order to the same order \cite{premnath2009incorporating,lycett2014multiphase}, thus the shift matrix $ {\bf{N}} $ is a lower triangular matrix.
As seen in Eq. (\ref{e10}), many combined terms are included in the orthogonal central moment set, which results in very tedious expressions in  $ {\bf{N}} $. In contrast, a non-orthogonal central moment set has been obtained in \cite{de2017nonorthogonal},
\begin{equation}\label{e11}
\begin{array}{l}
\left| {{{\tilde T}_i}} \right\rangle  = [{{\tilde k}_{000}},{{\tilde k}_{100}},{{\tilde k}_{010}},{{\tilde k}_{001}},{{\tilde k}_{110}},{{\tilde k}_{101}},{{\tilde k}_{011}},{{\tilde k}_{200}} - {{\tilde k}_{020}},{{\tilde k}_{200}} - {{\tilde k}_{002}},{{\tilde k}_{200}} + {{\tilde k}_{020}} + {{\tilde k}_{002}}, \\ 
{{\tilde k}_{120}} + {{\tilde k}_{102}},{{\tilde k}_{210}} + {{\tilde k}_{012}},{{\tilde k}_{201}} + {{\tilde k}_{021}},{{\tilde k}_{120}} - {{\tilde k}_{102}},{{\tilde k}_{210}} - {{\tilde k}_{012}},{{\tilde k}_{201}} - {{\tilde k}_{021}},{{\tilde k}_{111}},{{\tilde k}_{220}} + {{\tilde k}_{202}} + {{\tilde k}_{022}}, \\ 
{{\tilde k}_{220}} + {{\tilde k}_{202}} - {{\tilde k}_{022}},{{\tilde k}_{220}} - {{\tilde k}_{202}},{{\tilde k}_{211}},{{\tilde k}_{121}},{{\tilde k}_{112}},{{\tilde k}_{122}},{{\tilde k}_{212}},{{\tilde k}_{221}},{{\tilde k}_{222}}{]^ \top }.\\ 
\end{array}
\end{equation}
Compared with Eq. (\ref{e10}), the expressions in Eq. (\ref{e11}) is simplified to some extent, but some combined terms still exist and the corresponding $ \bf{N} $ is still tedious. In the present paper, we adopt the following central moment set,
\begin{equation}\label{e12}
\begin{array}{l} 
\left| {{{\tilde T}_i}} \right\rangle  = [{{\tilde k}_{000}},{{\tilde k}_{100}},{{\tilde k}_{010}},{{\tilde k}_{001}},{{\tilde k}_{110}},{{\tilde k}_{101}},{{\tilde k}_{011}},{{\tilde k}_{200}},{{\tilde k}_{020}},{{\tilde k}_{020}},{{\tilde k}_{120}},{{\tilde k}_{102}},{{\tilde k}_{210}}, \\ 
{{\tilde k}_{201}},{{\tilde k}_{012}},{{\tilde k}_{021}},{{\tilde k}_{111}},{{\tilde k}_{220}},{{\tilde k}_{202}},{{\tilde k}_{022}},{{\tilde k}_{211}},{{\tilde k}_{121}},{{\tilde k}_{112}},{{\tilde k}_{122}},{{\tilde k}_{212}},{{\tilde k}_{221}},{{\tilde k}_{222}}{]^ \top }. \\ 
\end{array}
\end{equation}
where the combined terms are completely eliminated. Correspondingly, the
raw moment set is $ \left| {{T_i}} \right\rangle  = [{k_{000}},{k_{100}},{k_{010}},\\
{k_{001}},{k_{110}},{k_{101}},{k_{011}},{k_{200}},{k_{020}},{k_{020}},{k_{120}},{k_{102}},{k_{210}},{k_{201}},{k_{012}},{k_{021}},{k_{111}},{k_{220}},{k_{202}},{k_{022}},{k_{211}},{k_{121}},{k_{112}},{k_{122}},{k_{212}},{k_{221}},\\{k_{222}}] $. However, it should be noted that the normal stress differences (${k_{200}}-{k_{020}}$ and ${k_{200}} - {k_{002}}$), and the trace of the pressure tensor ${k_{200}} + {k_{020}} + {k_{002}}$ should be considered separately in CLBM \cite{lycett2014binary}, so do the corresponding central moments ${{\tilde k}_{200}} - {{\tilde k}_{020}}$, ${{\tilde k}_{200}} - {{\tilde k}_{002}}$, and ${{\tilde k}_{200}} + {{\tilde k}_{020}} + {{\tilde k}_{002}}$. To handle this problem, the widely used diagonal relaxation matrix can be modified slightly to be a block-diagonal matrix (similar method has been previously used for 2D CLBM \cite{Asinari2008Generalized,fei2018modeling}),
\begin{equation}\label{e13}
{\bf{S}} = {\rm{diag}}\left\{ {{s_0},{s_1},{s_1},{s_1},{s_\nu },{s_\nu },{s_\nu },\left[ {\begin{array}{*{20}{c}}
		{{s_ + },{s_ - },{s_ - }}  \\
		{{s_ - },{s_ + },{s_ - }}  \\
		{{s_ - },{s_ - },{s_{\rm{ + }}}}  \\
		\end{array}} \right],{s_3},{s_3},{s_3},{s_3},{s_3},{s_3},{s_{3b}},{s_4},{s_4},{s_4},{s_{4b}},{s_{4b}},{s_{4b}},{s_5},{s_5},{s_5},{s_6}} \right\},
\end{equation}
with ${s_ + } = ({s_{2b}} + 2{s_2 })/3$ and ${s_- } = ({s_{2b}} - {s_2 })/3$. The kinematic and bulk viscosities are related to the relaxation parameters by $\nu  = (1/{s_2} - 0.5)c_s^2\Delta t$ and $ \xi  = 2/3(1/{s_{2b}} - 0.5)c_s^2\Delta t $, respectively. Thus the transformation matrix $ \bf{M} $ can be written explicitly according to Eqs. (\ref{e4}) and (\ref{e12}),
\begin{equation}\label{e14}
 {\bf{M} = }\left[ 
 \begin{array}{l r r r r r r r r r r r r r r r r r r r r r r r r r r}
1 &1 &1&1&1&1&1&1&1&1 &1 &1&1&1&1&1&1&1&1&1&1&1&1&1&1&1&1\\

0 &1 & -1&0&0&0&0&1&-1&1&-1&1&-1&1&-1&0&0&0&0&1&-1&1&-1&1&-1&1&-1\\
0 &0 &0&1&-1&0&0&1&1&-1 &-1 &0&0&0&0&1&-1&1&-1&1&1&-1&-1&1&1&-1&-1\\
0 &0 &0&0&0&1&-1&0&0&0&0 &1&1&-1&-1&1&1&-1&-1&1&1&1&1&-1&-1&-1&-1\\

0 &0 &0&0&0&0&0&1&-1&-1&1&0&0&0&0&0&0&0&0&1&-1&-1&1&1&-1&-1&1\\
0 &0 &0&0&0&0&0&0&0&0&0&1&-1&-1&1&0&0&0&0&1&-1&1&-1&-1&1&-1&1\\
0 &0 &0&0&0&0&0&0&0&0&0&0&0&0&0&1&-1&-1&1&1&1&-1&-1&-1&-1&1&1\\
0 &1 &1&0&0&0&0&1&1&1 &1 &1&1&1&1&0&0&0&0&1&1&1&1&1&1&1&1\\
0 &0 &0&1&1&0&0&1&1&1 &1 &0&0&0&0&1&1&1&1&1&1&1&1&1&1&1&1\\
0 &0 &0&0&0&1&1&0&0&0 &0 &1&1&1&1&1&1&1&1&1&1&1&1&1&1&1&1\\

0 &0 &0&0&0&0&0&1&-1&1 &-1 &0&0&0&0&0&0&0&0&1&-1&1&-1&1&-1&1&-1\\
0 &0 &0&0&0&0&0&0&0&0 &0 &1&-1&1 &-1&0&0&0&0&1&-1&1&-1&1&-1&1&-1\\
0 &0 &0&0&0&0&0&1&1&-1 &-1 &0&0&0&0&0&0&0&0&1&1&-1&-1&1&1&-1&-1\\
0 &0 &0&0&0&0&0&0&0&0 &0 &1&1&-1&-1&0&0&0&0&1&1&1&1&-1&-1&-1&-1\\
0 &0 &0&0&0&0&0&0&0&0 &0 &0&0&0&0&1&-1&1&-1&1&1&-1&-1&1&1&-1&-1\\
0 &0 &0&0&0&0&0&0&0&0 &0 &0&0&0&0&1&1&-1&-1&1&1&1&1&-1&-1&-1&-1\\

0 &0 &0&0&0&0&0&0&0&0 &0 &0&0&0&0&0&0&0&0&1&-1&-1&1&-1&1&1&-1\\

0 &0 &0&0&0&0&0&1&1&1 &1 &0&0&0&0&0&0&0&0&1&1&1&1&1&1&1&1\\
0 &0 &0&0&0&0&0&0&0&0&0 &1&1&1&1&0&0&0&0&1&1&1&1&1&1&1&1\\
0 &0 &0&0&0&0&0&0&0&0&0 &0&0&0&0&1&1&1&1&1&1&1&1&1&1&1&1\\
0 &0 &0&0&0&0&0&0&0&0&0 &0&0&0&0&0&0&0&0&1&1&-1&-1&-1&-1&1&1\\
0 &0 &0&0&0&0&0&0&0&0&0 &0&0&0&0&0&0&0&0&1&-1&1&-1&-1&1&-1&1\\
0 &0 &0&0&0&0&0&0&0&0&0 &0&0&0&0&0&0&0&0&1&-1&-1&1&1&-1&-1&1\\

0 &0 &0&0&0&0&0&0&0&0&0 &0&0&0&0&0&0&0&0&1&-1&1&-1&1&-1&1&-1\\
0 &0 &0&0&0&0&0&0&0&0&0 &0&0&0&0&0&0&0&0&1&1&-1&-1&1&1&-1&-1\\
0 &0 &0&0&0&0&0&0&0&0&0 &0&0&0&0&0&0&0&0&1&1&1&1&-1&-1&-1&-1\\

0 &0 &0&0&0&0&0&0&0&0&0 &0&0&0&0&0&0&0&0&1&1&1&1&1&1&1&1\\
 \end{array} \right].
 \end{equation}
 The shift matrix $ \bf{N} $ can be obtained analogously 
 according to the definition in Eq. (\ref{e4}).  Taking the fifth row of $ \bf{N} $, ${{\bf{N}}_4}$, as an example, the
 central moment ${{\tilde T}_4}$ can be expressed by raw moments from the lowest order to the same order,
\begin{equation}
 {{\tilde T}_4}{\rm{ = }}\left\langle {{f_i}\left| {({e_{ix}} - {u_x})({e_{iy}} - {u_y})} \right.} \right\rangle  = {u_x}{u_y}{T_0} - {u_y}{T_1} - {u_x}{T_2} + {T_4}.
\end{equation}
 Thus ${{\bf{N}}_4}$ is written as
\begin{equation}
 {{\bf{N}}_4} = [{u_x}{u_y}, - {u_y}, - {u_x},0,1,0,0,0,0,0,0,0,0,0,0,0,0,0,0,0,0,0,0,0,0,0,0].
\end{equation} 
The interested readers can refer to the Supplemental Material for the explicit expressions of ${{\bf{M}}}$, ${{\bf{N}}}$, ${{\bf{M}}^{ - 1}}$ and ${{\bf{N}}^{ - 1}}$.
 
 To implement the collision step in Eq. (\ref{e5}), the equilibria and forcing terms in central moment space, $\left| {\tilde T_i^{eq}} \right\rangle $ and $\left| {{C_i}} \right\rangle $, should be specified. In \cite{de2017nonorthogonal}, the discrete EDFs \cite{qian1992lattice}, $f_{D,i}^{eq} = {\omega _i}\rho [1 + ({{\bf{e}}_i}\cdot{\bf{u}})/c_s^2 + {({{\bf{e}}_i}\cdot{\bf{u}})^2}/2c_s^4 - {{\bf{u}}^2}/2c_s^2]$, are adopted. However, we argue three points against this choice: (1) it results in a lot of velocity terms in $\left| {\tilde T_i^{eq}} \right\rangle $ and $\left| {{C_i}} \right\rangle $, which is inconsistent with the 
physics of central moments; (2) it destroys the Galilean invariance for the off-diagonal elements of the third-order raw moments which are preserved naturally in the original CLBM by Geier \textit{et al.} \cite{geier2006cascaded}; (3) it leads to more computational cost for the calculation of $\left| {\tilde T_i^{eq}} \right\rangle $ and $\left| {{C_i}} \right\rangle $ compared with adopting the continuous equilibrium DF. In this work, $\left| {\tilde T_i^{eq}} \right\rangle $
are set equal to the continuous central moments of the continuous Maxwell-Boltzmann distribution \cite{geier2006cascaded,premnath2009incorporating,Asinari2008Generalized,lycett2014multiphase,lycett2014binary,fei2017consistent},
\begin{equation}\label{e15}
\left| {\tilde T_i^{eq}} \right\rangle  = {[\rho ,0,0,0,0,0,0,\rho c_s^2,\rho c_s^2,\rho c_s^2,0,0,0,0,0,0,0,\rho c_s^4,\rho c_s^4,\rho c_s^4,0,0,0,0,0,0,\rho c_s^6]^ \top}.
\end{equation}
The corresponding discrete EDF is in fact a generalized local equilibrium \cite{Asinari2008Generalized,premnath2009incorporating,fei2016thermal}. In addition, when the viscosity is quite small, i.e., $\nu  \sim O({10^{ - 7}})$, the higher than third order central moment equilibria can be also modified according to the factorized method \cite{premnath2011three,geier2009factorized}. Substituting the generalized local equilibrium into Eq. (\ref{e6}), we can get
\begin{equation}\label{e16}
\begin{aligned}
\left| {{C_i}} \right\rangle  &= {\bf{NM}}\left| {{R_i}} \right\rangle  \\
&={[0,{F_x},{F_y},{F_z},0,0,0,0,0,0,{F_x}c_s^2,{F_x}c_s^2,{F_y}c_s^2,{F_z}c_s^2,{F_y}c_s^2,{F_z}c_s^2,0,0,0,0,0,0,0,{F_x}c_s^4,{F_y}c_s^4,{F_z}c_s^4,0]^ \top}.
\end{aligned}
\end{equation}

We now summarize the computational algorithm of the above proposed implementation and forcing scheme in 3D CLBM: 

Step 1. Compute central moments $\left| {{{\tilde T}_i}} \right\rangle $ using the definition in Eq. (\ref{e2}). It should be noted that this method is quite basic and can be optimized by separating the transformation and shift sub-steps according to Eq. (\ref{e4}).

Step 2. Perform the collision step in Eq. (\ref{e5}).

Step 3. Reconstruct the post-collision raw moments according to $\left| {T_i^*} \right\rangle {\rm{ = }}{{\bf{N}}^{ - 1}}\left| {\tilde T_i^*} \right\rangle $,
and reconstruct the post-collision DFs according to $\left| {f_i^*} \right\rangle  = {{\bf{M}}^{ - 1}}\left| {T_i^*} \right\rangle $.

Step 4. Perform the streaming step and update the hydrodynamic variables according to Eqs. (\ref{e8}) and (\ref{e9}). Advance the time step and return to Step 1. 

It is known that the most computationally demanding part in 3D CLBM is the reconstruction step of the post-collision DFs \cite{geier2006cascaded,lycett2014binary,de2017nonorthogonal}.   
In the present work, the computational cost can be reduced significantly due to the facts that: firstly, the reconstruction step is divided into two sub-steps; secondly, the simplified central moment set is used such that the elements in ${{\bf{M}}^{ - 1}}$ and ${{\bf{N}}^{ - 1}}$ are much simplified.

In addition, the present method can also be used to simplify the 3D MRT-LBM. For example, it can be found that the non-orthogonal transformation matrix $ \textbf{M} $
in Eq. (\ref{e14}) and its inverse matrix ${{\bf{M}}^{ - 1}}$ have 343 and 216 non-zero elements, respectively. In  the orthogonal raw moment set by Premnath \textit{et al.} \cite{premnath2011three}, which has been used to construct the D3Q27 MRT-LBM in \cite{suga2015d3q27}, both the $ \textbf{M} $ and ${{\bf{M}}^{ - 1}}$ have 416 non-zero elements. Therefore, compared with the D3Q27 MRT-LBM in \cite{suga2015d3q27}, an MRT-LBM based on the present raw moment set can simplify the implementation and reduce the computational cost. As shown in  Appendix \ref{sec.5b}, the simulation results for 3D Lid-driven cavity flow by the non-orthogonal MRT-LBM are in good agreement with the benchmark solutions by Ku \textit{et al.} \cite{ku1987pseudospectral}, which implies that the non-orthogonal MRT-LBM can retain the numerical accuracy when simplifying the implementation.

\section{NUMERICAL SIMULATIONS}\label{sec.3}
In this section, several benchmark problems are conducted to verify the proposed 3D CLBM. In the simulation, the relaxation rates for the conserved central moments, ${s_0}$ and ${s_1}$ are set to be 1.0. Unless otherwise specified, the tunable relaxation parameters  for high-order central moments are also set to be 1.0. The standard half-way bounce-back boundary scheme is used for wall boundaries.
\subsection{The decay of a shear wave}\label{sec.3a}
Firstly, the decay of a shear wave on a moving frame is considered. The initial conditions of the flow are given as,
\begin{equation}\label{e17}
\rho (0) = 1.0, ~~{\bf{u}} = [A\sin (2\pi /L),0,B].
\end{equation}
where $A$ represents the initial amplitude of the shear wave, $ B $ is the reference velocity component, $L$ is the height of the computational domain. Periodic boundary conditions are used along the $ x $, $ y $ and $ z $ axes, and the analytical velocity field is ${u_x} = A\sin [\phi (y - Bt)]\exp ( - {\phi ^2}\nu t)$, where $\phi  = 2\pi /L$. For the simulations, the computational domain is covered by $5 \times 101 \times 5$ nodes. 

\begin{figure}
	\includegraphics[width=0.4\textwidth]{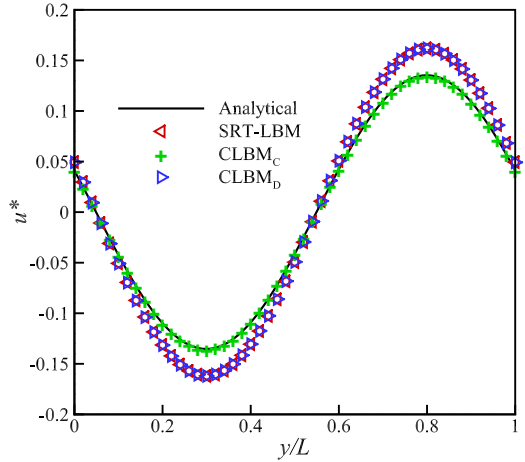}
	\caption{\label{FIG1a} Comparison of the velocity profiles for the decay of a shear wave at $ {t^*} = 2.0$ simulated by different methods.}
\end{figure}

\begin{figure}
	\includegraphics[width=0.4\textwidth]{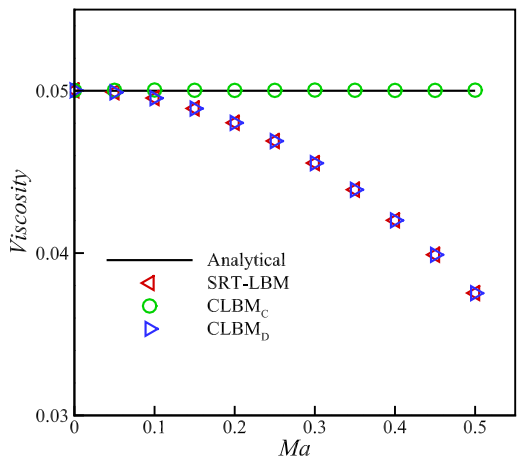}
	\caption{\label{FIG1b} Comparison of the simulated viscosities for the decay of a shear wave with $ Ma $ by different methods.}
\end{figure}

Firstly, we want to compare the central moment equilibria by Eq. (\ref{e15}) and by $f_{D,i}^{eq}$, which are denoted by ${\rm{CLB}}{{\rm{M}}_{\rm{C}}}$ and ${\rm{CLB}}{{\rm{M}}_{\rm{D}}}$, respectively. The SRT-LBM based on $f_{D,i}^{eq}$ is also used for comparison. The profiles for the dimensionless velocity $ {u^*} = u/A $ by different methods at the time $ {t^*} = {\phi ^2}\nu t = 2.0 $ and Mach number $ Ma = B/c_s^2 = 0.3 $ are shown in Fig.~\ref{FIG1a}. It is found that the simulation result by ${\rm{CLB}}{{\rm{M}}_{\rm{C}}}$ is in good agreement with the analytical solution, while there are visible differences between the numerical solutions by the other two methods and the analytical solution. 
Then, viscosity of the simulated fluid is obtained by measuring the time decay of the shear wave. The simulated viscosities at different $ Ma $ are compared in Fig.~\ref{FIG1b}, while the original imposed viscosity is  $\nu  = 0.05$. As is shown, the simulated viscosity by ${\rm{CLB}}{{\rm{M}}_{\rm{C}}}$ is independent of the reference velocity (or $ Ma $) and always agrees well with the imposed value. For the other two methods, the simulated viscosities decrease with the increase of the reference velocity. For example, the relative errors at 
$ Ma = 0.3 $ are $ 0.08\% $,  $ 8.92\% $, and $ 8.91\% $ for ${\rm{CLB}}{{\rm{M}}_{\rm{C}}}$, ${\rm{CLB}}{{\rm{M}}_{\rm{D}}}$, and SRT-LBM, respectively.

The above results confirm our argument in 
Sec. \ref{sec.2b} that using the discrete EDFs in CLBM as in \cite{de2017nonorthogonal} destroys the Galilean invariance (GI). It should be noted that there are two aspects to the issue of GI, one of which is related to the choice of the collision model and the other pertains to the choice of the discrete velocity model. For the standard lattice, the ${\rm{CLB}}{{\rm{M}}_{\rm{C}}}$ only preserves the GI naturally for the off-diagonal elements of the third-order raw moments. To restore the complete GI, additional correction terms \cite{prasianakis2009lattice} or more symmetrical lattice \cite{chikatamarla2006entropy} are needed.
 
\begin{table}
	\renewcommand\arraystretch{1.2}
	\caption{\label{TAB1}%
		Comparison of the CPU time (s) required by CLBM (${t_C} $) and SRT-LBM ($ {t_S} $) for different mesh sizes.
	}
	\begin{ruledtabular}
		\begin{tabular}{cccc}
			\textrm{ mesh }&
			\textrm{${t_C} $}&
			\textrm{$ {t_S} $}&
			\textrm{${t_C}/{t_L} $}\\
			\colrule
			$ 5 \times 101 \times 5 $ & 56.028 & 26.042 & 2.151\\
			
			$ 5 \times 201 \times 5 $  & 111.838 & 51.728 & 2.162\\
			$ 5 \times 401 \times 5 $  & 217.027 & 101.179 & 2.145\\
		\end{tabular}
	\end{ruledtabular}
\end{table}
We now verify the efficiency of the present simplified CLBM. We consider the same problem under different mesh sizes, i.e., $ 5 \times 101 \times 5 $, $ 5 \times 201 \times 5 $, and $ 5 \times 401 \times 5 $, and measure the CPU time required for 10000 iterations by the present CLBM and SRT-LBM. For each case, the CPU time is the average value after removing the minimum and maximum in nine runs. The code is developed based on C++, and runs on a laptop with Intel (R) Core (TM) i7-6500U CPU @ 2.5 GHz and RAM 8.00GB. The implementation is basic for both CLBM and SRT-LBM, without resorting to any optimization strategies, such as the preconditioning method \cite{hajabdollahi2017improving} and the LAPACK
library \cite{de2017nonorthogonal}. As shown in TABLE~\ref{TAB1}, the CPU time for the present 3D CLBM is around 2.15 times of the one by SRT-LBM. Compared with the method in \cite{de2017nonorthogonal}, where the computational cost overhead ratio is 13 to 14, the present implementation shows significant reduction in the computational cost.  

\subsection{Steady Poiseuille flow}\label{sec.3b}
The second problem considered is a steady Poiseuille flow driven by a constant body force $ {\bf{F}} = [{F_x},0,0] $. Thus the flow direction
is along the positive direction of the $ x $ axial. The analytical solution is $ {{\bf{u}}_a} = [{u_x},0,0] $, where ${u_x}(z) = {u_0}(1 - {z^2}/{L^2})$ and  $ L $ is the half-height of the channel. The peak velocity is ${u_0}{\rm{ = }}{F_x}{L^2}/2\nu $, by which the Reynolds number is defined as $ {\mathop{\rm Re}\nolimits}  = 2{u_0}L/\nu $. Due to the simple flow configuration, only 5 nodes are used to cover the length and width, and periodic boundary conditions are used in these two directions. 
\begin{figure}
	\includegraphics[width=0.4\textwidth]{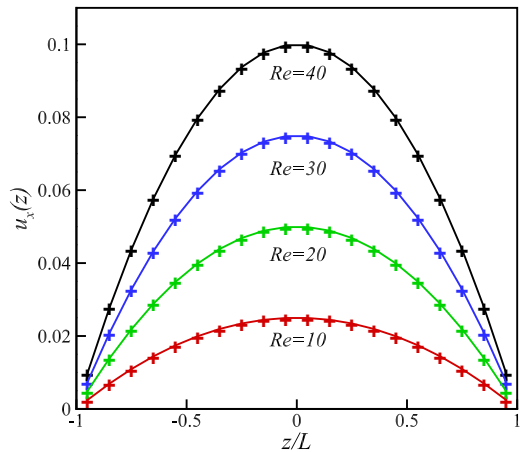}
	\caption{\label{FIG2a} Comparison between the numerical velocity profiles (symbols) and analytical solutions (solid lines) at $ {\mathop{\rm Re}\nolimits}  = [10,20,30,40] $ for the steady Poiseuille flow.}
\end{figure}
Firstly, we choose the kinematic viscosity $ \nu  = 0.1 $, and 20 nodes are employed to cover the channel height ($ 2L = 20\Delta x $). The velocity profiles at a series of Reynolds numbers, $ {\mathop{\rm Re}\nolimits}  = [10,20,30,40] $, are plotted in Fig.~\ref{FIG2a}. It can be seen in Fig.~\ref{FIG2a} that the simulation results are in very good agreement with the analytical solutions.
\begin{figure}
	\includegraphics[width=0.4\textwidth]{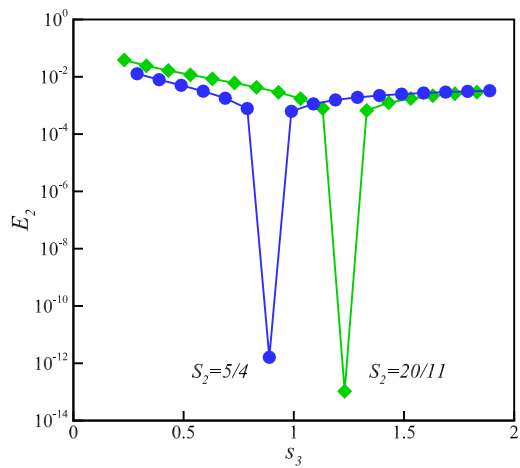}
	\caption{\label{FIG2b} Global relative error $ {E_2} $ changes with $ {s_3} $ for steady Poiseuille flow at $ \nu  = 0.1 $ and 0.2. When the nonslip rule $ {s_3} = (16 - 8{s_2})/(8 - {s_2}) $ is satisfied, $ {E_2} $ achieves a negligible value.}
\end{figure}
As analyzed  by previous studies \cite{ginzburg2003multireflection,Guo2008Analysis}, when the relaxation rate for the energy flux is set to be ${s_3} = (16 - 8{s_2})/(8 - {s_2})$, no numerical slip occurs in the steady Poiseuille flow for the MRT-LBM. It is shown recently in \cite{fei2017consistent} that among the existing forcing schemes in CLBM, only the consistent forcing scheme given in \cite{fei2017consistent} preserves the nonslip rule. Here we choose two cases with $ \nu  = 0.1 $ and 0.2 ($ {s_2} = 5/4 $ and 20/11), and measure the global relative errors at different ${s_3}$. The global relative error is defined as
\begin{equation}
{E_2} = \sqrt {\sum {{{({\bf{u}} - {{\bf{u}}_a})}^2}/\sum {{{\bf{u}}_a}^2}}},
\end{equation}
where the summation operator is over all grid nodes. As shown in Fig.~\ref{FIG2b}，
when $ {s_3} $ reaches the target value, $ {E_2} $ reduces significantly to a very small value, which confirms the consistent nonslip boundary condition in the present 3D CLBM.  
\begin{figure}
	\includegraphics[width=0.4\textwidth]{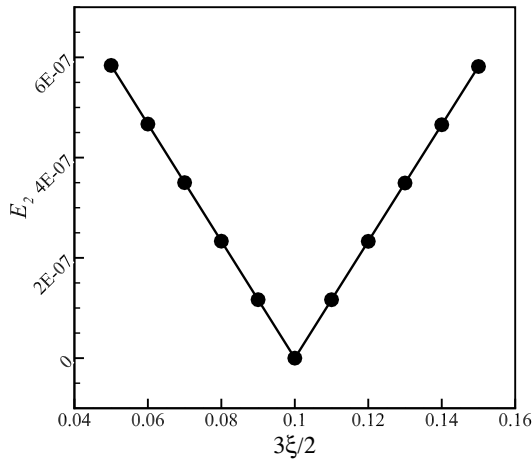}
	\caption{\label{FIG2c} Global relative error $ {E_2} $ changes with the
		bulk viscosity $ \xi $  for steady Poiseuille flow at $ \nu  = 0.1 $.}
\end{figure}
We now proceed to test the validity of the modification method in Eq. (\ref{e13}) to separately relax the stress differences and trace of the pressure tensor. As analyzed in \cite{lycett2014multiphase}, in the recovered Navier-Stokes equations by CLBM, the shear stress is expressed as $ \tau  = \rho \nu [\nabla {\bf{u}} + {\left( {\nabla {\bf{u}}} \right)^{\rm{T}}}] + \rho (\xi  - 2\nu /3)(\nabla \cdot{\bf{u}}){\bf{I}} $. As is known, the divergence-free condition cannot be accurately satisfied in LBM, thus the last term in ${\bf{\tau }}$ is an additional error for incompressible N-S equations and can be removed only when $\xi=2\nu/3$ (${s_{2b}} = {s_2}$). As shown in Fig.~\ref{FIG2c}, we choose $\nu=0.1$ coupled with the nonslip rule and measure $ {E_2} $ at different
$\xi$, and when $ \xi  = 2\nu/3 $, $ {E_2} $ reaches the smallest value. In addition, a linear relation between $ \left| {\xi  - 2\nu/3 } \right| $ and $ {E_2} $ can be also seen in Fig.~\ref{FIG2c}. Thus the validity of the modification method in Eq. (\ref{e13}) is confirmed.

\subsection{Steady flow through a square duct}\label{sec.3c}
\begin{figure}
	\includegraphics[width=0.45\textwidth]{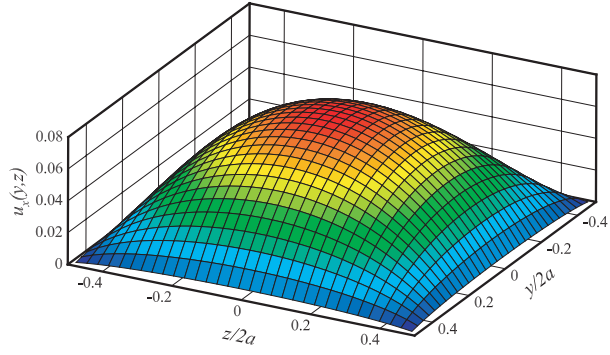}
	\caption{\label{FIG3a} Simulation result for the 
		velocity distribution of steady flow through a square duct. The simulation parameters are ${F_x} = 2 \times {10^{ - 4}}$, $ \nu=0.2 $, and
		$ a=16\Delta x$.}
\end{figure}

In this subsection, we consider the developed flow through a square duct where the flow field is variable in both the coordinate directions normal to the direction of the driving force $ {\bf{F}} = [{F_x},0,0]$ \cite{premnath2011three}. The flow has an analytical solution \cite{white2006viscous},
\begin{equation}
{u_x}(y,z) = \frac{{16{a^2}{F_x}}}{{\rho \nu {\pi ^3}}}\sum\limits_{n = 1,3,5,...}^{n = \infty } {{{( - 1)}^{(n - 1)/2}}\left[ {1 - \frac{{\cosh (n\pi z/2a)}}{{\cosh (n\pi /2)}}} \right]} \frac{{\cos (n\pi y/2a)}}{{{n^3}}},
\end{equation}
where $ a $ is the duct half-width, and $ - a \le y,z \le a$. In the simulation, we set ${F_x} = 2 \times {10^{ - 4}}$, $ \nu=0.2 $, and
$ a=16\Delta x$. The half-way bounce-back boundary condition is 
used at the walls, thus the grid lines are located at $ y = [ - 15.5\Delta x,..., - 0.5\Delta x,0.5\Delta x,...,15.5\Delta x] $ and $ z = [ - 15.5\Delta x,..., - 0.5\Delta x,0.5\Delta x,...,15.5\Delta x] $. Only 5 nodes are used to cover the $ x $ direction, along which the periodic boundary condition is adopted. The surface contour of the computed velocity field is shown in Fig.~\ref{FIG3a}. It is seen that the present CLBM can reproduce the velocity distribution for steady flow through the square duct, and the simulation result is in qualitative agreement with the analytical solution.
\begin{figure}
	\includegraphics[width=0.4\textwidth]{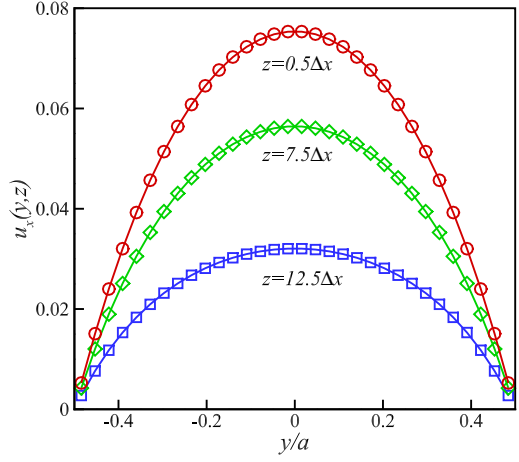}
	\caption{\label{FIG3b} Velocity profiles for steady flow through a square duct at $ z = [0.5\Delta x,7.5\Delta x,12.5\Delta x] $ (lines represent analytical solutions and symbols represent simulation results).}
\end{figure}
To be quantitative, the simulation results for the velocity profiles at $ z = [0.5\Delta x,7.5\Delta x,12.5\Delta x] $ are compared 
with the analytical solutions in Fig.~\ref{FIG3b}. It can be seen that the numerical results are in good agreement with the analytical solutions at different locations. In Sec. \ref{sec.3b}, we have found that the present CLBM holds the consistent nonslip boundary condition with the MRT-CLBM. It should be noted that the derivation of the nonslip rule  ${s_3} = (16 - 8{s_2})/(8 - {s_2})$ is based on the approximation
that the velocity field varies in only one coordinate direction \cite{Guo2008Analysis}. For the steady flow through a square duct, the velocity field varies in both $ y $ and $ z $ directions, thus the nonslip rule may be not applicable to this problem. To verify our argument, we define the local relative error over the $ z $ cross-section,
\begin{equation}
E(z) = \sqrt {\sum\nolimits_y {{{\left[ {{u_{nx}}(y,z) - {u_{ax}}(y,z)} \right]}^2}} /\sum\nolimits_y {{u_{ax}}{{(y,z)}^2}} }.
\end{equation}
where $ {{u_{ax}}} $ and $ {{u_{nx}}} $ denote the analytical and numerical velocities, respectively. Here we measure the change of $ E(z)$ with $ {s_3} $ at $z = [0.5\Delta x,7.5\Delta x,12.5\Delta x] $. As shown in Fig.~\ref{FIG3c},
when $ {s_3} $ reaches the target value 1.33077, $E(z)$ reduces to the smallest value for $ z = 0.5\Delta x $ and $ z = 7.5\Delta x $. However, the minimum point has a slight deviation from the target value for the case $ z = 12.5\Delta x $. The reason for the deviation is that the $ z = 12.5\Delta x $ cross-section is near to the wall boundary, where the flow contains 2D feature and the unidirectional approximation is destroyed. Although our argument is presented for the present CLBM, it should also apply to the MRT-LBM in general. As analyzed by Luo \textit{et al.} \cite{luo2011numerics} based on the 2D Lid-driven cavity flow simulation, the non-slip condition cannot be accurately satisfied for the complex flows, but usually a very good approximation can be obtained when using the non-slip rule.
\begin{figure}
	\includegraphics[width=0.4\textwidth]{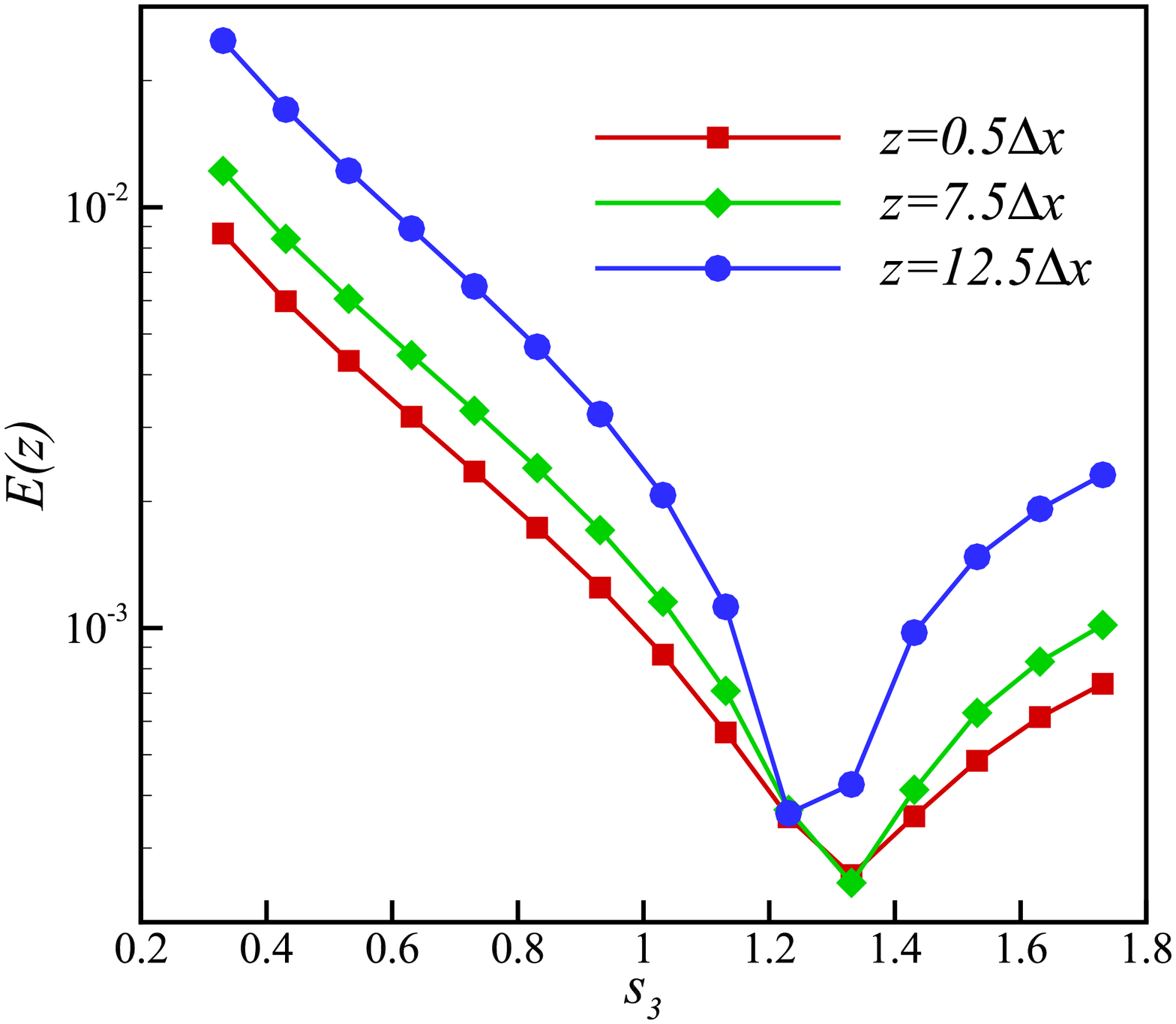}
	\caption{\label{FIG3c} Local relative error $ E(z)\ $ for
		steady flow through a square duct changes with $ {s_3} $ at $ z = [0.5\Delta x,7.5\Delta x,12.5\Delta x] $ (the kinematic viscosity is set to be $ \nu=0.2 $).}
\end{figure}
\subsection{Taylor-Green vortex flow}\label{sec.3d}
As a final example, we want to test the present CLBM with the consistent forcing scheme for an unsteady flow where the force field depends on both time and space. The considered problem is the Taylor-Green vortex flow \cite{taylor1937mechanism}, which has an analytical solution
\begin{eqnarray}
\begin{array}{l}
{u_x} =  - {u_0}\cos ({k_1}x)\sin ({k_2}y)\exp [ - \nu (k_1^2 + k_2^2)t],\\
{u_y} = {u_0}\frac{{{k_1}}}{{{k_2}}}\sin ({k_1}x)\cos ({k_2}y)\exp [ - \nu (k_1^2 + k_2^2)t].
\end{array}
\end{eqnarray}
where $ {u_0} $ is the amplitude of the imposed velocity field, and 
$ {k_1} $ and $ {k_2} $ denote the wave numbers along the $ x $ and $ y $
directions. The body force is given by 
\begin{equation}
\begin{array}{l}
{F_x} =  - (\rho {k_1}u_0^2/2)\sin (2{k_1}x)\exp [ - 2\nu (k_1^2 + k_2^2)t],\\
{F_y} =  - (\rho k_1^2u_0^2/2{k_2})\sin (2{k_2}y)\exp [ - 2\nu (k_1^2 + k_2^2)t].
\end{array}
\end{equation}
In the simulation, the computational domain is defined in $ 0 \le x,y \le 2\pi $, and covered by $ L \times L  $ grid points, while only 5 nodes are adopted along the $ z $ direction. Thus the wave numbers are $ {k_1} = {k_2} = 2\pi /L $. To eliminate the compressibility effect, $ {u_0} $ is set to be 0.005. 
\begin{figure}
	\includegraphics[width=0.4\textwidth]{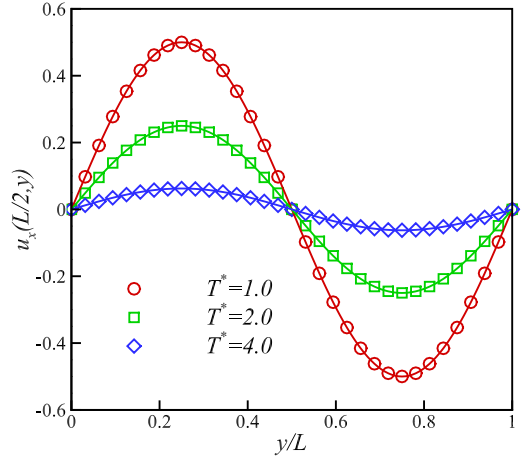}\\
	(a)\\
	\includegraphics[width=0.4\textwidth]{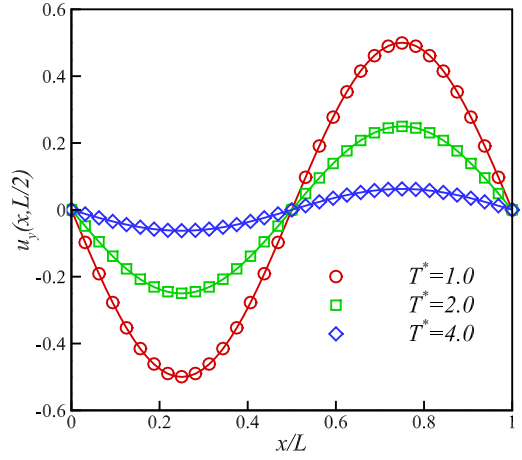}\\
	(b)	
	\caption{\label{FIG4a} Numerical results (symbols) and analytical solutions (solid lines) of the Taylor-Green vortex flow at $ Re=50 $
		and $ {T^*} = [1.0,2.0,4.0] $: (a) horizontal velocity profile in the  $ x = L/2 $ cross-section, (b) vertical velocity profile in the $ y = L/2 $ cross-section.}
\end{figure}
\begin{figure}
	\includegraphics[width=0.4\textwidth]{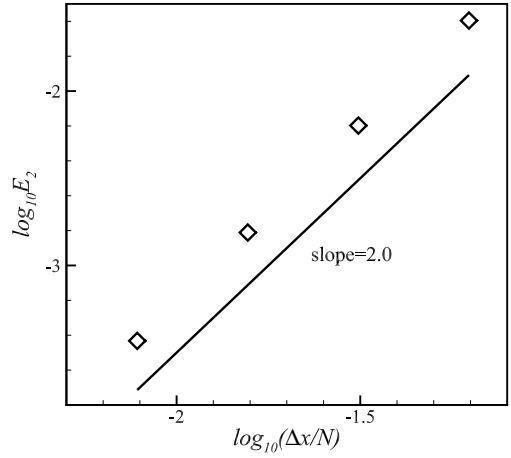}
	\caption{\label{FIG4b} Global relative error $ {E_2} $ as a function of the grid spacing for the Taylor-Green vortex flow at $ Re=50 $ and $ {T^*} = 4.0 $. Symbols represent the measured global relative errors, and the fit
		slope of the results is 2.0345.}
\end{figure}
Firstly, we choose $L = 32\Delta x $  and $ \nu = 0.0032 $, and the corresponding Reynolds number $ {\mathop{\rm Re}\nolimits}  = {u_0}L/\nu $ is 50. The numerical results of the horizontal velocity profile in the  $ x = L/2 $ cross-section and the vertical velocity profile in the $ y = L/2 $ cross-section are plotted in Fig. ~\ref{FIG4a}, from which it can be seen that the numerical results are in good agreement with the analytical solutions at different non-dimensional time, ${T^*} =\nu (k_1^2 + k_2^2)t/\ln 2$. Then simulations characterized by a series of grid resolutions are carried out,
$ L/\Delta x = [16,32,64,128]$, while the kinematic viscosity is obtained through $ \nu  = {u_0}L/{\mathop{\rm Re}\nolimits} $. The log-log plot of the 
global relative error $ {E_2} $ at $ {T^*} = 4.0 $  as a function of the grid spacing is presented in Fig.~\ref{FIG4b}, and the fit slope of the numerical results is 2.0345. This demonstrates that the present CLBM with the consistent forcing scheme has second-order accuracy in space.
\section{CONCLUSIONS}\label{sec.4}
In this work, we present an efficient 3D CLBM formulation with a consistent forcing scheme for a general force field.
In the method, the most computationally demanding reconstruction step in 3D CLBM is divided into two sub-steps based on the GMRT framework. A very simple central moment set is adopted to construct the cascaded operator such that the tedious combined terms in the shift matrix $ \textbf{N} $ are completely eliminated. To match the separate relaxations for certain second-order moments, the previously used  diagonal relaxation matrix is modified to be a block-diagonal matrix.

Our proposed method is very efficient and easy to implement. Through the simulation of the decay of a shear wave, it is confirmed the present method can significantly reduce the computational cost compared with the recently proposed non-orthogonal CLBM \cite{de2017nonorthogonal}. The inconsistency  of adopting the discrete EDF in the non-orthogonal CLBM \cite{de2017nonorthogonal} is  revealed and discussed. Then the consistent forcing scheme is verified by simulating several benchmark force-driven flows, highlighting very good properties of the developed methodology in terms of accuracy, convergence and consistency with the nonslip rule.

The implementation method developed is quite intelligible and general. Although the D3Q27 lattice is adopted for the derivation, the corresponding implementation in other lattices can be readily obtained. The derivation method puts the 3D MRT-LBM and CLBM into a unified general framework. Thus the developed methodology is not only applicable to CLBM but can also be adopted to simplify the 3D MRT-LBM, as demonstrated in Appendix B.

\begin{acknowledgments}
Support from the MOST National Key Research and Development Programme (Project No. 2016YFB0600805) and the Center for Combustion Energy at Tsinghua University is gratefully acknowledged. Supercomputing time on ARCHER is provided by the “UK Consortium on Mesoscale Engineering Sciences (UKCOMES)” under the UK Engineering and Physical Sciences Research Council Grant No. EP/L00030X/1.
\end{acknowledgments}

\appendix
\section{D3Q19 CLBM}\label{sec.5a}
For the D3Q19 lattice, the discrete velocities $ {{\bf{e}}_i} = \left[ {\left| {{e_{ix}}} \right\rangle ,\left| {{e_{iy}}} \right\rangle ,\left| {{e_{iz}}} \right\rangle } \right] $ $ (i=0,1,...,18) $ are defined as 
\begin{equation}\label{ea1}
\begin{array}{l}
\left| {{e_{ix}}} \right\rangle  = {[0,1, - 1,0,0,0,0,1, - 1,1, - 1,1, - 1,1, - 1,0,0,0,0]^ \top }, \\ 
\left| {{e_{iy}}} \right\rangle  = {[0,0,0,1, - 1,0,0,1,1, - 1, - 1,0,0,0,0,1, - 1,1, - 1]^ \top }, \\ 
\left| {{e_{i{\rm{z}}}}} \right\rangle  = {[0,0,0,0,0,1, - 1,0,0,0,0,1,1, - 1, - 1,1,1, - 1, - 1]^ \top }. \\ 
\end{array} 
\end{equation}
The central moment set can be extracted from Eq. (\ref{e12}) 
\begin{equation}\label{ea2}
\begin{array}{l} 
\left| {{{\tilde T}_i}} \right\rangle  = {[{{\tilde k}_{000}},{{\tilde k}_{100}},{{\tilde k}_{010}},{{\tilde k}_{001}},{{\tilde k}_{110}},{{\tilde k}_{101}},{{\tilde k}_{011}},{{\tilde k}_{200}},{{\tilde k}_{020}},{{\tilde k}_{020}},{{\tilde k}_{120}},{{\tilde k}_{102}},{{\tilde k}_{210}},{{\tilde k}_{201}},{{\tilde k}_{012}},{{\tilde k}_{021}},{{\tilde k}_{220}},{{\tilde k}_{202}},{{\tilde k}_{022}},]^ \top }. 
\end{array}
\end{equation}
The block-diagonal matrix is given by 
\begin{equation}\label{ea3}
{\bf{S}} = {\rm{diag}}\left\{ {{s_0},{s_1},{s_1},{s_1},{s_\nu },{s_\nu },{s_\nu },\left[ {\begin{array}{*{20}{c}}
		{{s_ + },{s_ - },{s_ - }}  \\
		{{s_ - },{s_ + },{s_ - }}  \\
		{{s_ - },{s_ - },{s_{\rm{ + }}}}  \\
		\end{array}} \right],{s_3},{s_3},{s_3},{s_3},{s_3},{s_3},{s_4},{s_4},{s_4}} \right\}.
\end{equation}
The transformation matrix $ \textbf{M} $ can be written explicitly according to
Eqs. (\ref{e4}) and (\ref{ea2})
\begin{equation}\label{ea4}
{\bf{M} = }\left[ 
\begin{array}{l r r r r r r r r r r r r r r r r r r r r r r r r r r}
1 &1 &1&1&1&1&1&1&1&1 &1 &1&1&1&1&1&1&1&1\\

0 &1 & -1&0&0&0&0&1&-1&1&-1&1&-1&1&-1&0&0&0&0\\
0 &0 &0&1&-1&0&0&1&1&-1 &-1 &0&0&0&0&1&-1&1&-1\\
0 &0 &0&0&0&1&-1&0&0&0&0 &1&1&-1&-1&1&1&-1&-1\\

0 &0 &0&0&0&0&0&1&-1&-1&1&0&0&0&0&0&0&0&0\\
0 &0 &0&0&0&0&0&0&0&0&0&1&-1&-1&1&0&0&0&0\\
0 &0 &0&0&0&0&0&0&0&0&0&0&0&0&0&1&-1&-1&1\\
0 &1 &1&0&0&0&0&1&1&1 &1 &1&1&1&1&0&0&0&0\\
0 &0 &0&1&1&0&0&1&1&1 &1 &0&0&0&0&1&1&1&1\\
0 &0 &0&0&0&1&1&0&0&0 &0 &1&1&1&1&1&1&1&1\\

0 &0 &0&0&0&0&0&1&-1&1 &-1 &0&0&0&0&0&0&0&0\\
0 &0 &0&0&0&0&0&0&0&0 &0 &1&-1&1 &-1&0&0&0&0\\
0 &0 &0&0&0&0&0&1&1&-1 &-1 &0&0&0&0&0&0&0&0\\
0 &0 &0&0&0&0&0&0&0&0 &0 &1&1&-1&-1&0&0&0&0\\
0 &0 &0&0&0&0&0&0&0&0 &0 &0&0&0&0&1&-1&1&-1\\
0 &0 &0&0&0&0&0&0&0&0 &0 &0&0&0&0&1&1&-1&-1\\

0 &0 &0&0&0&0&0&1&1&1 &1 &0&0&0&0&0&0&0&0\\
0 &0 &0&0&0&0&0&0&0&0&0 &1&1&1&1&0&0&0&0\\
0 &0 &0&0&0&0&0&0&0&0&0 &0&0&0&0&1&1&1&1\\
\end{array} \right].
\end{equation}
Thus the equilibria and forcing terms in the central moment space can be given as 
\begin{equation}\label{ea5}
\left| {\tilde T_i^{eq}} \right\rangle  = {[\rho ,0,0,0,0,0,0,\rho c_s^2,\rho c_s^2,\rho c_s^2,0,0,0,0,0,0,\rho c_s^4,\rho c_s^4,\rho c_s^4,]^ \top}.
\end{equation}
and 
\begin{equation}\label{ea6}
\begin{aligned}
\left| {{C_i}} \right\rangle  = {[0,{F_x},{F_y},{F_z},0,0,0,0,0,0,{F_x}c_s^2,{F_x}c_s^2,{F_y}c_s^2,{F_z}c_s^2,{F_y}c_s^2,{F_z}c_s^2,0,0,0]^ \top}.
\end{aligned}
\end{equation}
The explicit expressions of ${{\bf{M}}}$, ${{\bf{N}}}$, ${{\bf{M}}^{ - 1}}$ and ${{\bf{N}}^{ - 1}}$ for the D3Q19 CLBM are also provided in the Supplemental Material.

\section{Non-orthogonal MRT-LBM for 3D Lid-driven cavity flow}\label{sec.5b}
\begin{figure}
	\includegraphics[width=0.4\textwidth]{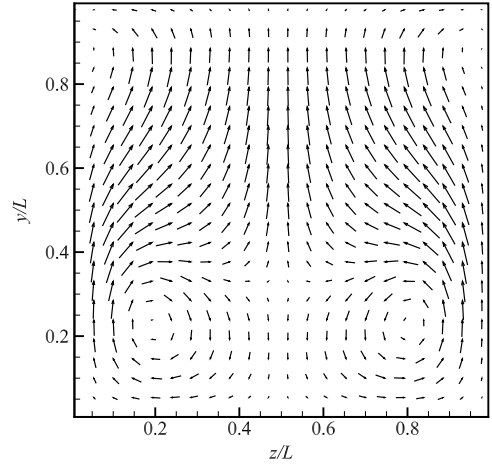}
	\caption{\label{FIG5a} The flow direction vector at 
		${\mathop{\rm Re}\nolimits}  = 500$ in the $x = 0.5L$ plane.}
\end{figure}
In the D3Q27 non-orthogonal MRT-LBM, the raw moment equilibria are obtained by $\left| {T_i^{eq}} \right\rangle  = {\bf{M}}\left| {f_{D,i}^{eq}} \right\rangle $. Although the force 
field is not considered in this case, the forcing terms in raw moment space can be directly obtained by $\left| {G_i^{eq}} \right\rangle  = {\bf{M}}\left| {{R_{G,i}}} \right\rangle $, where
${R_{G,i}} = {\omega _i}[({{\bf{e}}_i} - {\bf{u}})/c_s^2 + ({{\bf{e}}_i}\cdot{\bf{u}}){{\bf{e}}_i}/c_s^4]{\bf{F}}$ refer to the forcing scheme by Guo \textit{et al.} \cite{guo2002discrete}.
Analogously, a D3Q19 non-orthogonal MRT-LBM can be constructed using the transformation matrix ${{\bf{M}}}$ in Eq. (\ref{ea4}).
\begin{figure}
	\includegraphics[width=0.4\textwidth]{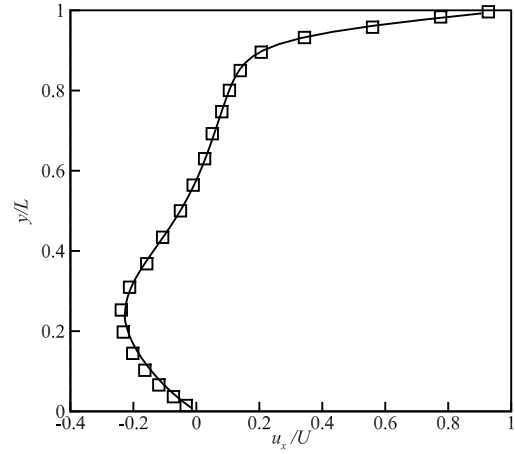}\\
	(a)\\
	\includegraphics[width=0.4\textwidth]{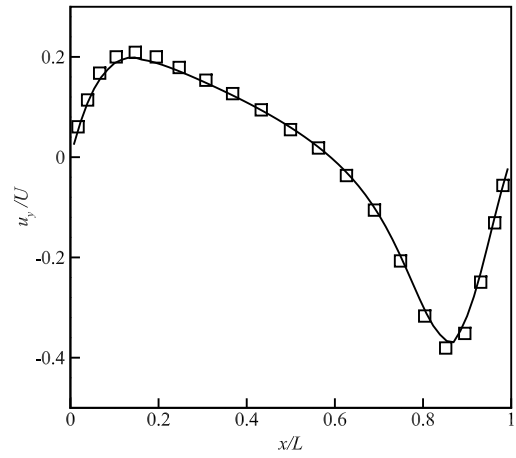}\\
	(b)	
	\caption{\label{FIG5b} The non-orthogonal MRT-LBM simulations (solid lines) and benchmark solutions \cite{ku1987pseudospectral} (symbols) of the 3D Lid-driven cavity flow at $ Re=400 $: (a) horizontal velocity profiles in the  vertical centerline, (b) vertical velocity profile in the horizontal centerline.}
\end{figure}
We now use the D3Q27 non-orthogonal MRT-LBM to simulate the 3D Lid-driven cavity. The flow is confined in a cubic box 
$L \times L \times L$ and driven by a top lid at $y = L$ with constant velocity $U = 0.1$. The Reynolds number ${\mathop{\rm Re}\nolimits}  = UL/\nu =400$ is considered in the simulation, and the length of the cubic box is set to be $L = 64\Delta x$. From Fig.~\ref{FIG5a}, it can be seen that a pair of vortices are located near the bottom of the $x = 0.5L$ plane, which is consistent with the results reported in \cite{ku1987pseudospectral}. In addition, the velocity profiles by the non-orthogonal MRT-LBM are compared with the benchmark solutions \cite{ku1987pseudospectral} in Fig.~\ref{FIG5b}. It can be seen the present simulation results are in good agreement with the benchmark solutions. Furthermore, it is found that the non-orthogonal MRT-LBM  requires approximately 25\% less computational time than the orthogonal MRT model in \cite{suga2015d3q27}.
\nocite{*}

\bibliography{3DCLBM}

\end{document}